\documentclass[letterpaper, 10 pt, conference]{ieeeconf}  % Comment this line out if you need a4paper

\IEEEoverridecommandlockouts                              % This command is only needed if 
\usepackage{cite}
\usepackage{amsmath,amssymb,amsfonts}
\usepackage{algorithmic}
\usepackage{graphicx}
\usepackage{textcomp}
\usepackage{xcolor}
\usepackage{multirow}
\usepackage{tabularx}
\usepackage{booktabs}

\title{\LARGE \bf
Field Deployment of Multi-Agent Reinforcement Learning Based Variable Speed Limit Controllers
}

\author{Yuhang Zhang$^{\dagger}$, Zhiyao Zhang$^{\dagger}$, Marcos Qui\~{n}ones-Grueiro$^{\dagger}$, William Barbour$^{\dagger}$, \\ Clay Weston$^{\ddagger}$, Gautam Biswas$^{\dagger}$, Daniel Work$^{\dagger}$
\thanks{
$^{\dagger}$Institute for Software Integrated Systems, Vanderbilt University, Nashville TN, 37212, USA. $^{\ddagger}$Southwest Research Institute, San Antonio, TX, 78238, USA.
}
\thanks{
The authors are grateful to  Caliper for technical support on the TransModeler micro-simulation software used in this work. The authors would like to thank the Tennessee Department of Transportation (TDOT), Southwest Research Institute (SwRI), Arcadis, and Stantec, who assisted in this deployment, and Caleb Van Geffen and Josh Scherer from Vanderbilt University for their development work of AI-DSS. The contents of this report reflect the views of the authors, who are responsible for the facts and accuracy of the information presented herein. This work is supported by a grant from the U.S. Department of Transportation Grant Number 693JJ22140000Z44ATNREG3202. This material is based upon work supported by the National Science Foundation under Grant No. CNS-2135579 (Work). The U.S. Government assumes no liability for the contents or use thereof. 
}
}

\begin{document}

\maketitle
\thispagestyle{empty}
\pagestyle{empty}

%%%%%%%%%%%%%%%%%%%%%%%%%%%%%%%%%%%%%%%%%%%%%%%%%%%%%%%%%%%%%%%%%%%%%%%%%%%%%%%%
\begin{abstract}
This article presents the first field deployment of a multi-agent reinforcement-learning (MARL) based variable speed limit (VSL) control system on the I-24 freeway near Nashville, Tennessee. We describe how we train MARL agents in a traffic simulator and directly deploy the simulation-based policy on a 17-mile stretch of Interstate 24 with 67 VSL controllers. We use invalid action masking and several safety guards to ensure the posted speed limits satisfy the real-world constraints from the traffic management center and the Tennessee Department of Transportation. Since the time of launch of the system through April, 2024, the system has made approximately 10,000,000 decisions on 8,000,000 trips. The analysis of the controller shows that the MARL policy takes control for up to 98\% of the time without intervention from safety guards. The time-space diagrams of traffic speed and control commands illustrate how the algorithm behaves during rush hour. Finally, we quantify the domain mismatch between the simulation and real-world data and demonstrate the robustness of the MARL policy to this mismatch.
\end{abstract}

%%%%%%%%%%%%%%%%%%%%%%%%%%%%%%%%%%%%%%%%%%%%%%%%%%%%%%%%%%%%%%%%%%%%%%%%%%%%%%%%
\section{Introduction}
% \subsection{Motivation}

% 1. Infrustructure based traffic control
% 2. Variable speed limit
% 3. RL applications
Infrastructure-based traffic control systems have been pivotal in managing road traffic long before the advent of vehicle-based control technologies. These systems, including traffic signal control, ramp metering, and variable speed limit (VSL), form the backbone of efforts to streamline traffic flow and enhance safety on roadways. VSL is unique in that it controls the mainline freeway flow by altering speed limits in response to real-time traffic conditions, thus aiming to reduce congestion and accidents~\cite{lu2014review, khondaker2015variable}.

Historically, most deployed VSL control systems have been rule-based~\cite{elefteriadou2012variable, robinson2000examples}. These systems dynamically adjust speed limits based on predefined traffic characteristics, such as flow and density thresholds. The simplicity of rule-based systems contributes to their widespread adoption, as they do not require complex computational resources or extensive training data. However, such simplicity can be a limitation; they may not adapt to unforeseen traffic scenarios or optimize for multiple conflicting objectives, such as minimizing travel time while maximizing safety.

\begin{figure}
    \centering
    \includegraphics[width=1\columnwidth]{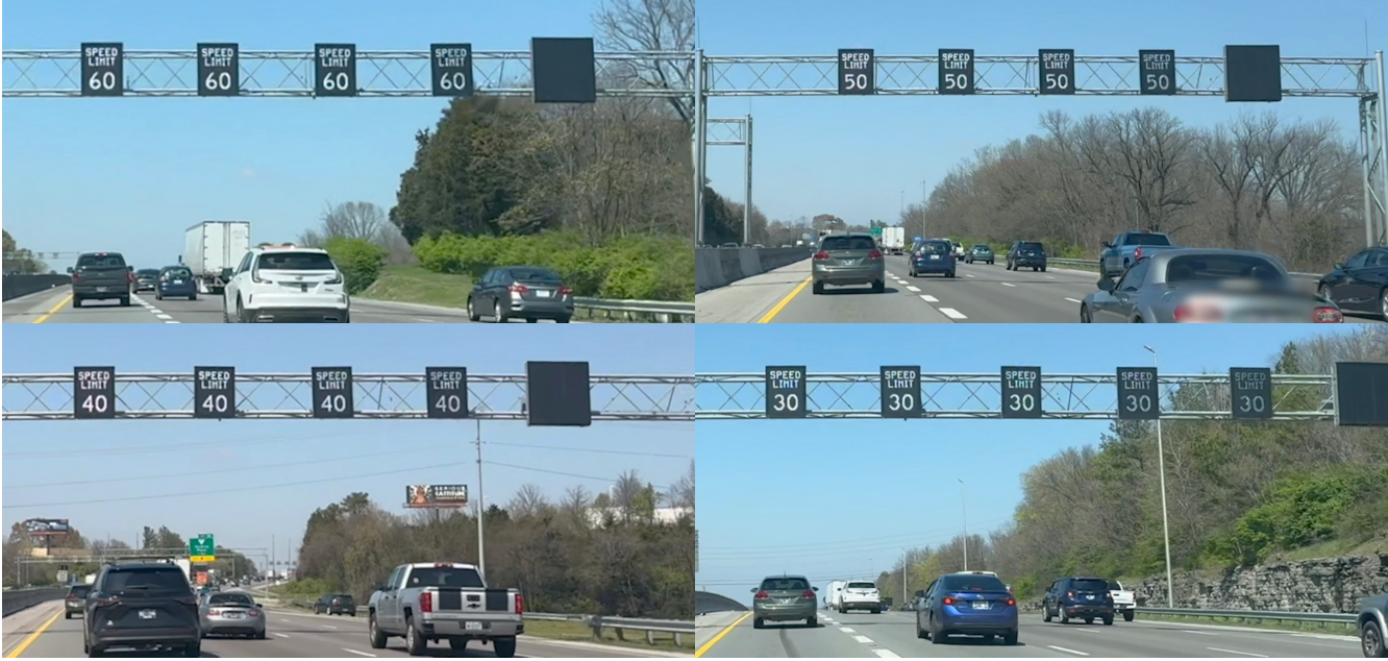}
    \caption{The MARL-based VSL control system on I-24 Westbound: This figure shows a consecutive four gantries from a driver's perspective when approaching a congestion tail. As drivers proceed, they encounter progressively reduced speed limits of 60, 50, 40, and 30 mph displayed on each gantry, sequentially alerting them to the upcoming slow-down pattern.}
    \label{fig:vsl gantries}
\end{figure}

% \IEEEpubidadjcol
\textit{Reinforcement learning} (RL) is an emerging approach for decision and control in variety of applications ranging from strategic game playing, industry robotics and complex decision-making~\cite{sutton2018reinforcement}. Within the realm of traffic control, the ability of RL to learn and adapt from interaction with an environment makes it potentially promising for managing the dynamic and often unpredictable nature of road traffic. 

Earlier studies have applied RL to VSL control in simulated environments, demonstrating its potential to outperform traditional methods by adapting to evolving traffic conditions and optimizing for multiple objectives simultaneously~\cite{kuvsic2020overview}. These results are promising, yet the transition from simulated environments to real-world applications is unexplored. This gap represents a critical barrier: while simulation offers a controlled setting to fine-tune algorithms, real-world traffic presents additional complexities such as varying driver behavior, diverse vehicle types, and unpredictable weather conditions, all of which can affect the performance of RL-based strategies. Consequently, real deployments can offer further insights about the potential of RL to work in operational traffic management systems. % \Zhiyao{The sim-to-real gaps are not evaluated yet: it remains unclear that how the RL policy would be impacted when real-world traffic dynamics, e.g., varying driver behavior and diverse vehicle types (not to mention weather since we don't evaluate it in bad weather), add unsimulated complexities.} Valid, not sure what to do about it at this stage.

% \subsection{Problem Statement and Contribution}
%The main problem addressed in this work is following: \textit{RL is promising, but how can we deploy RL-based VSL controllers in real world and evaluate the performance?}

% \Zhiyao{Contribution: We close the loop from simulation training to field deployment of RL for infrastructure-based highway traffic control.}

\begin{figure*}
    \centering
    \includegraphics[width=2\columnwidth]{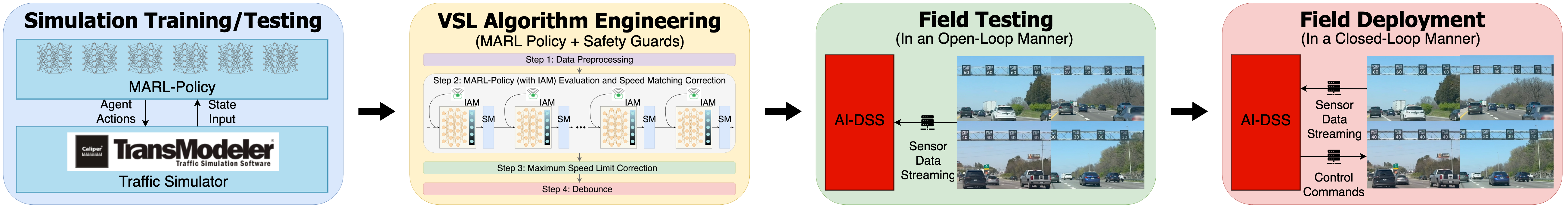}
    \caption{Deployment pipeline of our MARL-based VSL: \textit{\textbf{Step 1:}} We trained 8 agents in TransModeler on a 7-mile stretch of I-24 and then tested it with 34 agents on a 17-mile stretch of westbound I-24 with various simulation parameters. \textit{\textbf{Step 2:}} We extracted the optimal policy learned from simulation and applied invalid action masking and safety guards to satisfy real-world constraints. \textit{\textbf{Step 3:}} We tested the behavior of the proposed MARL-based VSL control algorithm in an open-loop manner, with continuous streaming of I-24 sensor data feeding into Artificial-Intelligence Decision Support System (AI-DSS), the infrastructure software served for communication with Traffic Management Center (TMC). \textit{\textbf{Step 4:}} We deployed the MARL-based VSL control algorithm in a closed-loop manner across 67 VSL gantries spanning a 17-mile stretch of I-24 with nearly 160,000 daily commuters on March 8, 2024.} %\Zhiyao{Actually this could be the overview picture of this paper. I still insist on moving Figure 1 to the end}
    \label{fig:pipeline}
\end{figure*}

The main contribution of this work is to describe and provide a preliminary assessment of the first field deployment of a \textit{multi-agent reinforcement learning} (MARL) based VSL control system encompassing 67 VSL gantries on a 17-mile (each direction) segment of Interstate-24 (I-24) near Nashville, Tennessee, USA (Figure~\ref{fig:vsl gantries}). Figure~\ref{fig:pipeline} overviews the deployment pipeline of our MARL-based VSL controllers. Specifically, our contributions are as follows:
\begin{itemize}
    % \item We train and test a MARL-based policy in a simulated environment with domain randomization to enhance its generalizability to unknown compliance rates and traffic demand.
    \item We train a MARL-based policy in a simulated environment where homogeneous agents are exposed to diverse traffic scenarios. The optimal policy, once derived, is subsequently tested in a different simulated environment with variable system parameters to assess its robustness and adaptability.
    \item We refine the optimal policy derived from simulation by incorporating invalid action masking and several safety guards designed to meet real-world constraints. 
    % \Zhiyao{We build safety filters that incorporate operational constraints and traffic authority requirements to filter out invalid RL actions}
    \item We deploy the MARL-based VSL control algorithm in the field.  Evaluation results indicate that the MARL-based policy autonomously makes up to 98\% of the final decisions without any intervention from the safety guards. %\Zhiyao{domain mismatch?}
\end{itemize}

Since its initial deployment on March 8, 2024, the MARL-based VSL control system has operated continuously, making decisions at 30-second intervals, 24 hours a day. It has generated over 10 million control decisions, impacting more than 8 million trips through the corridor. It continues to operate today.

% \subsection{Outline of the Work}
The remainder of the article is organized as follows. Section~\ref{sec:literature} reviews the related works on VSL field deployments and RL-based controller design. Section~\ref{sec:methods} presents our processes to train in simulation and deploy on the live I-24 VSL system. Section~\ref{sec:experiment} describes the setup of I-24 where the VSL controllers are deployed. Section~\ref{sec:results} provides the preliminary results of the deployment. Section~\ref{sec:conclusions} concludes the paper and provides the future directions.

%%%%%%%%%%%%%%%%%%%%%%%%%%%%%%%%%%%%%%%%%%%%%%%%%%%%%%%%%%%%%%%%%%%%%%%%%%%%%%%%
\section{Related Work}
\label{sec:literature}
\subsection{VSL Field Deployments}
VSL systems were first proposed and deployed in the 1960's. Since then, various VSL control systems have been implemented across Europe, Australia, New Zealand, and North America \cite{lu2014review, han2009best, robinson2000examples, papageorgiou2008effects}. These deployments have demonstrated benefits in enhancing traffic safety and homogenizing traffic. For instance, a study in Belgium observed an 18\% reduction in injury crashes and a 20\% reduction in rear-end collisions following the implementation of VSL \cite{de2018safety}.

Most VSL systems employ rule-based control algorithms, where speed limits are dynamically adjusted based on predefined thresholds related to traffic characteristics. Due to its simplicity, this approach has been widely adopted in numerous deployments \cite{elefteriadou2012variable, zhang2022quantifying}. Although model-based control algorithms have been proposed \cite{carlson2010optimal, yu2018optimal}, only a few have undergone empirical validation in real-world settings. Notably, the SPECIALIST algorithm, which is based on shock wave theory, demonstrated in simulations the capability to reduce travel times. Subsequently, it was implemented on a 14 km segment of the A12 freeway in the Netherlands, resolving shock waves in nearly 80\% of cases when activated \cite{hegyi2008specialist, hegyi2010dynamic}. Another instance is the implementation of a model predictive control (MPC) based VSL algorithm on Whitemud Drive in Edmonton, Canada, where preliminary results indicated improved average travel speeds \cite{wang2016implementation}.

\subsection{RL-based VSL Control}
% first RL-based VSL control, model-based RL to improve sample efficiency, MARL for VSL

Over the past decade, RL-based VSL control algorithm design has gained significant attention within the traffic community \cite{kuvsic2020overview} due to its ability to manage complex dynamic systems. The authors in \cite{7949069} proposed a Q-learning based algorithm to enhance traffic efficiency, training a single VSL controller in a simulation setting. A lane-dependent VSL approach was explored in~\cite{wu2020differential}, where the authors evaluated an actor-critic based algorithm with various reward designs, including travel time, safety, and pollution. 
% To overcome the sample inefficiency of conventional model-free RL algorithms, the work in \cite{jin2024variable} introduced and validated a freeway tunnel VSL control strategy based on a model-based RL framework.

In recent years, MARL has proven to be an effective approach for control in multi-agent systems. The study \cite{wang2019new} developed a cooperative VSL control system using distributed RL within a \textit{vehicle-to-infrastructure} (V2I) environment to optimize freeway traffic mobility and safety, significantly reducing total travel time and speed variances between freeway segments, which indicates a lower risk of rear-end collisions. In \cite{zheng2023coordinated}, the authors employed the MADDPG algorithm across four VSL controllers in a network with consecutive bottlenecks, designing a reward function to maintain bottleneck density below critical levels to avoid capacity drops and enhance traffic flow. Moreover, our previous work \cite{zhang2023marvel} introduced MARVEL, a MARL framework designed for large-scale VSL control across extensive freeway corridors. The policy derived from the MARVEL framework exhibited superior traffic mobility performance compared to baseline algorithms and demonstrated generalizability across varying traffic networks, demands, and compliance rates.

The above-mentioned methods have demonstrated remarkable performance in traffic simulators; however, none have been deployed on real freeway systems to validate their effectiveness.

\subsection{RL-based Field Experiments in Transportation}
% Eugene et al. deployed car paper ICRA, Kathy Jang paper (scaled test bed + circles RL deployment paper), Yu Han survey paper.
% In this section, we aim to summarize some existing field experiments of RL-based controllers in the transportation field. 
Most existing RL-based field experiments have focused on \textit{connected and automated vehicles} (CAVs) due to their notable potential to stabilize traffic flow and reduce energy consumption \cite{stern2018dissipation}. Jang et al. \cite{jang2019simulation} conducted zero-shot policy transfer experiments on a scaled testbed, finding that a policy with noise injected into the state and action space could achieve a 5\% reduction in travel time in a roundabout scenario, compared to a policy without noise injection. Chalaki et al. \cite{chalaki2020zero} extended this method by integrating adversarial learning during training, demonstrating that the adversarially trained policy outperforms the Gaussian noise injection approach.

Lichtlé et al. \cite{lichtle2022deploying} developed a pipeline that bypasses the tedious calibration of simulators by using real-world trajectory data to directly learn controllers. They successfully deployed their controller on actual vehicles in freeway traffic, highlighting its potential for energy savings. In a landmark study in November 2022, Jang et al. \cite{jang2024reinforcement} deployed RL-based controllers on 100 vehicles driving on I-24 in Nashville, marking the largest field test of automated vehicles aimed at smoothing traffic flow.

%%%%%%%%%%%%%%%%%%%%%%%%%%%%%%%%%%%%%%%%%%%%%%%%%%%%%%%%%%%%%%%%%%%%%%%%%%%%%%%%
\section{Methods} \label{sec:methods}
% \Zhiyao{My concern is the Intro and Related Work take more than the first two pages. The manuscript already reaches the page limit without the conclusion section. I feel like the related work is a bit too long for a conference paper. }
In this section, we briefly review how we formulate the VSL control into a MARL problem and train the MARL policy in a microscopic traffic simulator as discussed in our previous works~\cite{10207650,zhang2023marvel}. Moreover, to guarantee real-world constraints from the transportation agency, we apply invalid action masking on MARL policy and introduce safety guards to override certain actions. Lastly, we demonstrate the final implemented control algorithm by detailing each step.
% how we approach the final implementation by integrating all components.   

\subsection{Problem Formulation}
% Here we describe how we formulate the large-scale VSL control problem into a MARL problem and the corresponding agent, state, action and reward design.
We consider a large-scale VSL control system where multiple VSL controllers span a long freeway segment with nearly evenly distributed distance. We formulate this problem into a cooperative MARL problem, which can be modeled as a Markov Game, defined as a tuple $\langle\{\mathcal{S}^i\}_{i\in \{1,\dots, n\}}, \{\mathcal{A}^i\}_{i\in \{1,\dots, n\}}, \{\mathcal{R}^i\}_{i\in \{1,\dots, n\}}, P, n, \gamma\rangle$ for a total of $n$ agents, where $\mathcal{S}^i$ denotes the local state space for agent $i$, $\mathcal{A}^i$ denotes the action space for agent $i$, $\mathcal{R}^i$: $\{\mathcal{S}^i\}_{i\in \{1,\dots, n\}} \times \{\mathcal{A}^i\}_{i\in \{1,\dots, n\}} \times \{\mathcal{S}^i\}_{i\in \{1,\dots, n\}}\rightarrow \mathbb{R}$  denotes the reward for agent $i$, $P$: $\{\mathcal{S}^i\}_{i\in \{1,\dots, n\}} \times \{\mathcal{A}^i\}_{i\in \{1,\dots, n\}} \times \{\mathcal{S}^i\}_{i\in \{1,\dots, n\}}\rightarrow [0, 1]$ denotes the transition probability of the environment from a given state to the next state. The goal of MARL for each agent is to learn a policy that maximizes its own cumulative discounted reward:
\begin{align}
    J^i(\theta_1, \dots, \theta_n)=\mathbb{E}_{S_t, A_t}\left[\sum_{t=0}^T \gamma^t r_t^i \right],
\end{align}
where $S_t$ denotes the global state concatenating all local states at time $t$ 
% \Zhiyao{I think it might be better to denote it like: $S_t$ is the global state as time $t$ and the local state for each agent $i$ as $s^i_t \in S_t$ is a subset of $S_t$} 
% \Zhiyao{So we can make the hierarchy consistent: mathcal for unified space, lowercase for elements in the unified space, and capital letters for a product space of all elements at a time}
, $A_t=(a_t^1, \dots, a_t^n)$ denotes the joint action of all agents at time $t$ and $r_t^i$ the reward of agent $i$ at time $t$.
For the MARL components, we adopt the following system: 

\textbf{Agent:} each VSL controller on a highway gantry is represented by an agent. To improve the scalability of the system, we consider the agents as homogeneous and they share the same parameters.

\textbf{State Space:} $s^i_t = \langle a_t^{i-1}, \nu_t^i, o_t^i, \nu_t^{i+1}, o_t^{i+1} \rangle$, where $a_t^{i-1}$ is the closest downstream agent's intended action at time $t$, and $\nu_t^i, o_t^i, \nu_t^{i+1}, o_t^{i+1}$ the average traffic speed, the average traffic occupancy from traffic sensor assigned to the agent $i$ and the closest upstream agent $i+1$. All these input features are normalized to $[0, 1]$. We assume $a_t^{i-1}$ is the default maximum speed limit for $i=1$ (the most downstream agent). 
% \Zhiyao{directly say $a_t^{0}$ for $i=1$?}

\textbf{Action Space:} $A^i = \langle30, 40, 50, 60, 70 \rangle$, which is a set of speed limit values that satisfy field deployment requirements. 
% \Zhiyao{Here actions are both space and set. I'd prefer define actions in a set since it's not structured}

\textbf{Reward:} the reward function encompasses three terms, namely, adaptability, safety, and mobility. The adaptability term is used to penalize an agent posting high-speed limits when the traffic is in congestion and is used to help the agent to identify the congestion state. The safety term encourages the agents to coordinate with each other to generate a slow-down speed profile upstream of the congestion tail. The mobility term encourages the agents to post a higher speed limit when traffic condition allows. Finally, the reward function for agent $i$ at time $t$ is the following:
\begin{align}
    r_t^i&=w_a r_{t}^{i,a} + w_s r_{t}^{i,s} + w_m r_{t}^{i,m}
\end{align}
where $r_{t}^{i,a}, r_{t}^{i,s}, r_{t}^{i,m}$ represents the adaptability, safety and mobility terms, respectively, and $w_a, w_s, w_m$ represent the corresponding coefficients.  For more details on the structure and design of the reward function, please refer to our previous work~\cite{zhang2023marvel}.%\Zhiyao{probably just write down the math definition of the three terms and refer readers to the previous work for design principles of the reward. It's important to deliver clearly the definition but here we only tell people `why the rewards'} \Dan{I suggest the opposite, cite the prior paper for the details; leave the reasoning in for this one.}

%^In summary, by integrating three reward terms, the agents are encouraged to first identify the congestion state and post the minimum speed limit, then generate a slow-down speed profile for upstream drivers to reduce speed variation, and eventually improve mobility as much as possible. 

\subsection{Training and Testing in Simulation}
% \Dan{this section needs some work, it's hard to follow and is a wall of text. you'll see i'm adding in paragraphs to make it easier to structure. }
% This section discusses the specific configurations of the training and testing traffic scenarios, including the geometry network, traffic demand, and compliance rate.

We use the microscopic simulation software TransModeler for all simulations used to train our VSL controllers.
% \Zhiyao{Reference \cite{yu2022the} justifies why we use MAPPO and \cite{zhang2023marvel} is our previous work on the algorithm. They deliver separate information so let's separate them. Also can we put MARVEL reference at the begging of this section so it's clear that the algorithmic part is not new and we designed it previously.} 
Transmodeler allows driver compliance with the regulatory VSL system to be modeled. We set the compliance rate of 5\% for the training scenarios as we expect the compliance rate on the freeway to be relatively low.  

 We train our policy using the Multi-Agent Proximal Policy Optimization (MAPPO) algorithm~\cite{yu2022the}. The training scenario is a seven-mile long freeway segment with four lanes on I-24 westbound in Nashville, USA. We implement eight VSL controllers at half-mile intervals upstream of an on-ramp merging area aimed at learning a cooperative policy with varying traffic conditions. A traffic sensor is co-located with each VSL controller to capture the traffic characteristics, with data updated every minute. 
% \Dan{OK but this is a bit odd, the real world is using 30 sec data and the sim is using 1 min data?}\Yuhang{initially we were using 1 min data for the real world and changed that to 30s later} \Dan{cool just add one sentence in the deployment when you mention that the data rate is 30 sec, which is shorter than we assumed in training as the data rates changed as the system evolved. }

To induce traffic congestion, we set a single two-lane on-ramp merging area with a flow around 1000 veh/lane/hr.
% \Dan{I'm confused. are there 2 lanes for the on-ramp? How many on-ramps are there? I'm getting a detail but not the big picture. maybe above you can rephrase to "a single 2-lane on-ramp merging area" which will disambiguate }
The simulation spans two hours, during which the mainstream inflow is initially set at 1850 veh/lane/hr for the first hour to induce congestion. For the second hour, this rate is reduced to half to alleviate the congestion. These variations in traffic speed and traffic density create a variety of traffic conditions, offering the homogeneous agents an extensive range of scenarios to navigate. 

We test the learned policy on a 17-mile segment of I-24 in simulation, which replicates half of the targeted field network. We focus on the westbound traffic encompassing 34 VSL controllers with one traffic sensor placed 0 to 0.2 miles downstream of each VSL controller, replicating the real conditions on I-24.  We consider three testing scenarios including multiple congestion and various compliance rates. Our previous results in~\cite{zhang2023marvel} demonstrate that the learned policy is able to scale to a greater number of VSL controllers and generalize to new environments with different traffic settings from the training scenario. The traffic scenarios under the control of the learned policy exhibit a superior mobility performance compared to a state-of-the-practice control algorithm that was initially deployed on I-24, while maintaining a lower speed variation to improve safety. 

\subsection{Real-World Constraints}
% \Zhiyao{I assume this subsection is what the safety guards are about, but this phrase is never mentioned here.}
In this section, we detail the real-world constraints pertinent to the intended deployment of the VSL control algorithm, along with our proposed solutions to ensure that our final control algorithm meets these criteria. 
% \Dan{you need to reference your figures. here talk about fig 3. you also need to say the order in which you apply the logic. you show it in the figure but don't say anything about it in the text.}\Yuhang{I put the integration part in section D, maybe we want to have an overview here?}
\subsubsection{Maximum Step-Down Constraint}
The Manual on Uniform Traffic Control Devices (MUTCD) specifies a maximum permissible speed limit differential of 10 miles per hour (mph) between each pair of VSL controllers that are part of a group indicating slowdown traffic patterns~\cite{MUTCD2009}. For instance, pointing at the downstream of traffic, a sequence of speed limits set at $[70, 60, 50]$ mph complies with the regulation but  $[70, 50, 30]$ mph does not. Our safety reward term is designed to promote satisfaction of this constraint but we may still violate it during testing~\cite{zhang2023marvel}.

To ensure adherence to this constraint, we implement a technique known as invalid action masking (IAM)~\cite{huang2020closer}. This technique introduces a masking layer following the output of the policy network during the testing and deployment period, which effectively removes invalid actions. It thereby restricts the sampling process to the subset of valid actions, ensuring compliance with the specified speed limit differential. We define the invalid action set of agent $i$ at time $t$ according to the following equation:
\begin{align}
    I=\{a | a > a_t^{i-1} + a_{\text{diff}}\}
\end{align}
where $a_t^{i-1}$ is the closest downstream agent's intended action at time $t$, $a_{\text{diff}}$ is the maximum permissible speed limit differential for slowing down, which is $10$ mph.

\subsubsection{Speed-Matching Constraint}
As an operating requirement from the transportation authority, the posted speed limits should not significantly deviate from actual traffic speeds. This requirement allows the speed limits to be easily explained to motorists, even if it prevents more exotic wave dissipation designs from being implemented.

\textit{Proposed Safety Guard:} To align with this requirement, we implemented a mapping function applied to certain outputs generated by the learned policy. This function is defined as follows:
\begin{align}
    V=
    \begin{cases}
        \text{clip}(30, \text{min}(a_t^{i-1}+a_{\text{diff}}, f(\nu_t^i)), 70)
        &
        \text{if $a_t^i = 30$} \\
        \text{clip}(30, f(\nu_t^i), 70)
        &
        \hspace{-50pt}\text{if $a_t^i=70$ and $o_t^i\geq o_{\text{thred}}$}
    \end{cases}
    \label{mappingfunction}
\end{align}
where $\text{clip}(a,\cdot,b)$ is a clip function with minimum bound $a$ and maximum bound $b$, $f(\cdot)$ is a mapping function to map the input to the nearest multiple of 10 that is greater than the input, $o_{\text{thred}}$ is occupancy threshold, which is used to determine whether to apply this mapping when agents are selecting 70 mph.  
% \Zhiyao{can we have a subscript for V? Also how is $V'$ used? Is there a hierarchy/order among these rules? Because $V$ appears in speed matching and max speed limit, it looks unclear whether the V in max speed limit is from speed matching output. Also why do we have the capital V instead of lowercases?}\Yuhang{I changed the order of them and now these rules are in a sequence as seen from figure 3}

% \Yuhang{fill in here}In summary, we apply formula~\ref{mappingfunction} only to 

\begin{figure}
    \centering
    \includegraphics[width=1\columnwidth]{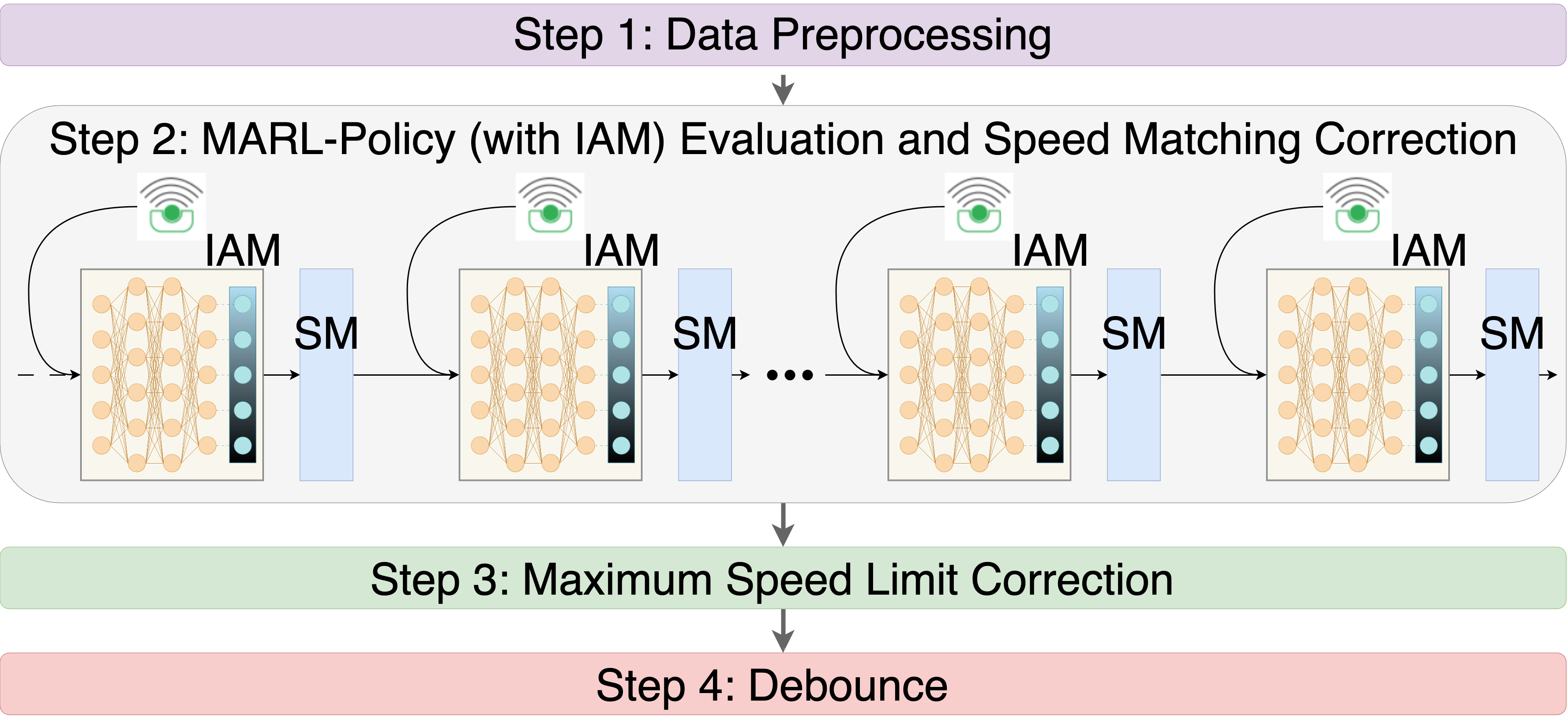}
    \caption{The deployed VSL control algorithm, centered around a MARL policy, considers all real-world constraints. IAM represents ``Invalid Action Masking'' and SM represents ``Speed-Matching''.}
    \label{fig:algorithm workflow}
\end{figure}

\subsubsection{Maximum Speed-Limit Constraint}
The maximum allowable speed limit on a freeway segment is determined by several factors, including geometric design and safety considerations. Consequently, maximum speed limits may vary across different segments. Currently, the majority of segments within the targeted field network are subject to a maximum speed limit of 70 mph, while others are capped at 65 or 55 mph. %It is important to note that the Tennessee Department of Transportation (TDOT) may request adjustments to these maximum speed limits in the future, based on the findings of new engineering studies.

\textit{Proposed Safety Guard:} To ensure adherence to this constraint while maintaining homogeneous MARL settings for scalability, we apply a clip function to assure the posted speed limit is within the allowable range. Specifically, for any generated speed limit $V$, we apply the following equation:
\begin{align}
    V'=\text{min}(V,V_{\text{max}})
    \label{formula:maximum_constraint}
\end{align}
where $V_{\text{max}}$ is the allowable maximum speed limit and $V'$ is the clipped speed limit that satisfies this constraint.

\begin{figure*}
    \centering
    \includegraphics[width=2\columnwidth]{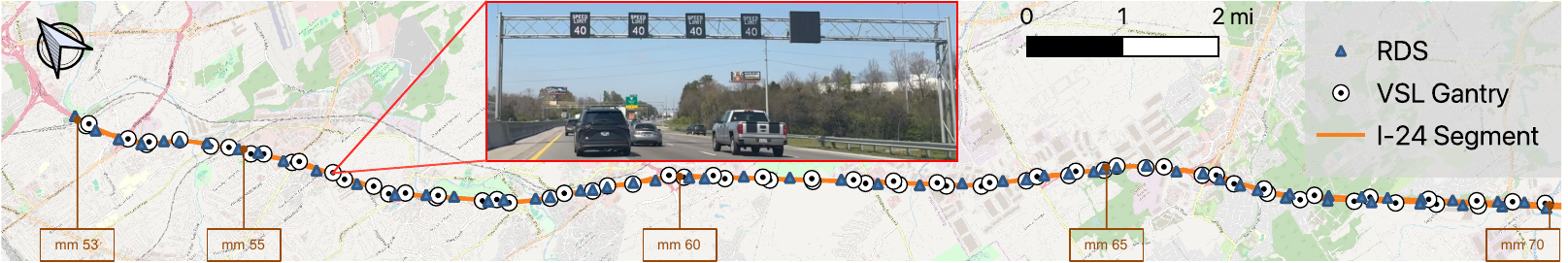}
    \caption{Overview of the VSL deployment segment of both directions on I-24 SMART Corridor. The left direction is going downtown Nashville and right is going to Murfreesboro. RDS denotes Radar Detection System, which is the traffic sensor installed on I-24.}
    \label{fig:smartcorridor}
\end{figure*}

\subsubsection{Debounce Constraint}
A \textit{bounce} is defined as a spatial sequence of speed limits at the same timestamp where all intermediate speed limits are higher than both the first and last speed limits in the sequence, which are referred to as boundary speed limits. The \textit{order} of a bounce is defined as the number of intermediate speed limits in the bounce sequence. For instance, within the direction of traffic flow, a sequence of $[30, 60, 50]$ is a bounce with order $1$ while a sequence of $[30, 60, 50, 40]$ is a bounce with order $2$. As per local design requirements, the deployed algorithm should not generate any bounce with order $1$.

\textit{Proposed Safety Guard:} To align with this requirement, we iterate all intended speed limits and identify every bounce with order of $1$. We apply the following equation to override the intermediate speed limit of each identified bounce:
\begin{align}
    V''=\text{min}(V_d', V_u')
    \label{formula:debounce}
\end{align}
where $V_d', V_u'$ are the two boundary speed limits and $V''$ is the corrected speed limit that satisfies debounce constraint.

\subsection{Algorithm Integration}
In this section, we explain the general pipeline of the deployed algorithm, from data preprocessing to the final outputs. The architecture of the deployed algorithm for a set of gantries in one direction of travel is shown in Figure~\ref{fig:algorithm workflow}. This algorithm has four steps as follows:
\begin{itemize}
    \item Step 1: Process all sensor data to interpolate missing values and to determine the critical downstream sensor for each VSL controller. This critical sensor will be used to provide state inputs in Step 2.
    \item Step 2: For each VSL controller, evaluate the MARL policy with all state inputs as described in Section~\ref{sec:methods}. With invalid action masking, the output of the policy network ensures the maximum step-down constraint. This output will go through the speed-matching module for any necessary adjustments. The updated output will then be used as a part of the state inputs to feed the upstream VSL controllers. The VSL controllers are processed in order starting with the most downstream gantry first, and the output of this step is a set of initial speed limits that are corrected in later steps. This step is responsible for satisfying the maximum step-down and speed-matching constraints.
    \item Step 3: Process all VSL controllers (starting from the most downstream gantry) to make maximum speed limit corrections according to~\eqref{formula:maximum_constraint}. This step is responsible for satisfying maximum speed limit constraint.
    % \Dan{I fixed these. don't refer to formulas or equations. use eqref and that's it}.
    \item Step 4: Process all VSL controllers again (starting from the most downstream gantry) to identify if any debounce constraints are violated, and correct them with the debounce logic in~\eqref{formula:debounce} to generate the final speed limits to be posted. This step is responsible for satisfying debounce constraint.
\end{itemize} 

\section{Experimental Setup} \label{sec:experiment}
In this section, we provide a detailed overview of the deployment, which is known as the I-24 SMART Corridor. We also describe the software infrastructure, which is known as the \textit{Artificial Intelligence Decision Support System} (AI-DSS). The AI-DSS supports the implementation of our MARL-based VSL control algorithm, amongst other decision support functionalities not described in this work.

\begin{figure*}
    \centering
    \includegraphics[width=2\columnwidth]{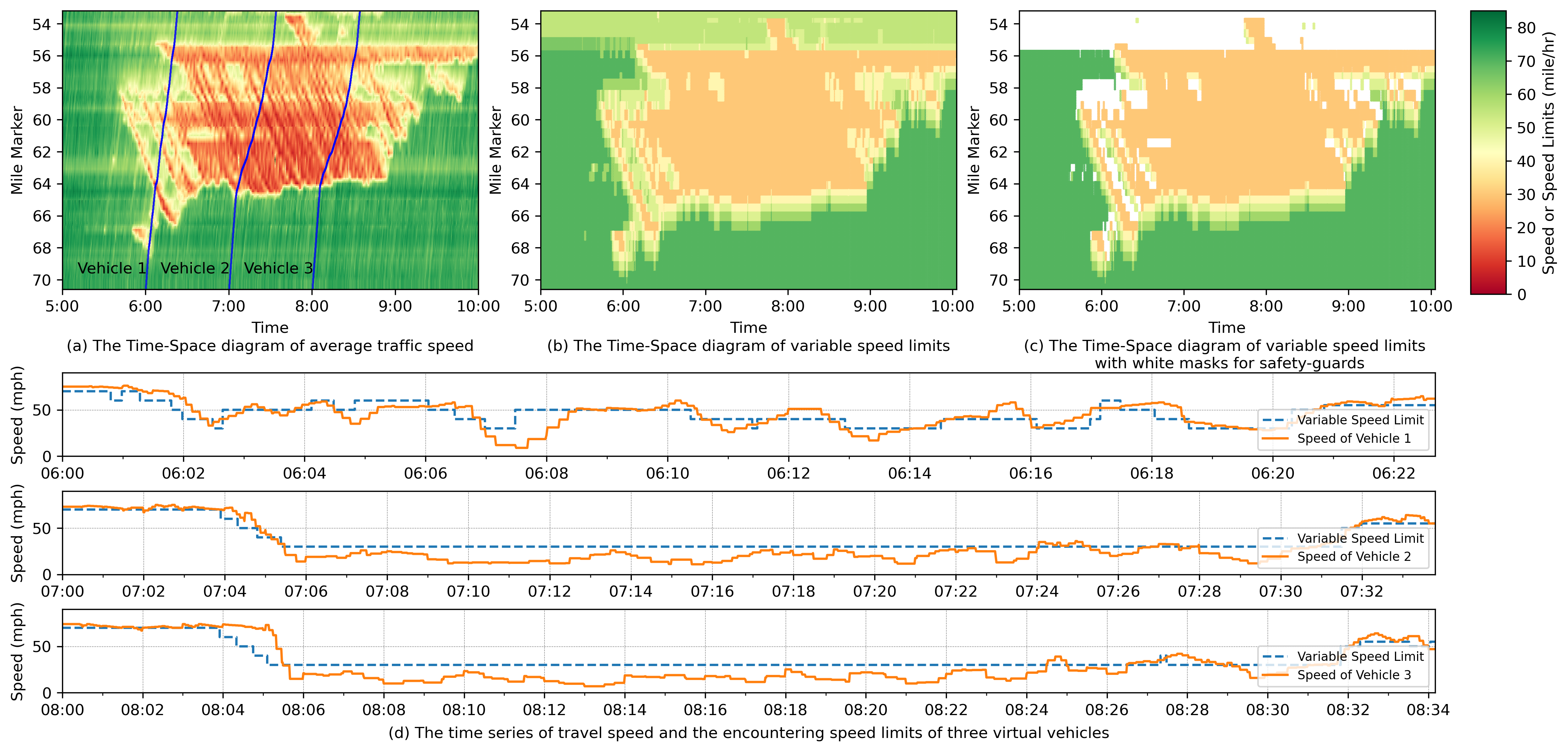}
    \caption{The MARL-based VSL control algorithm's behavior from a random morning peak hour (Monday, April 22, 2024) on I-24 Westbound: (a) displays the time-space diagram of average traffic speed recorded by roadside RDS sensor in every 30 seconds. With x-axis representing time and y-axis representing mile markers, the traffic direction is going upward along y-axis to Nashville. Three virtual vehicles are simulated starting from 6am, 7am and 8am, according to the RDS speed data, and their trajectories are overlayed on the figure. Controlling at every 30-second interval, (b) presents the time-space diagram of the 34 VSL gantries controlled by the MARL-based algorithm in this study. (c) shows the same diagram as (b) but with safety guards overrides masked as white. (d) details the time series of the travel speed and the encountering speed limits of each virtual vehicle generated in (a).}
    \label{fig:control algorithm behavior}
\end{figure*}

\subsection{I-24 SMART Corridor} 
% 1. location, network geometry, Murfreesboro (Rutherford County), Nashville (Davidson County)
% 2. traffic characteristics
% 3. the first ICM project in Tennessee
The I-24 SMART Corridor is the first \textit{Integrated Corridor Management} (ICM) project in Tennessee, and it includes a set of strategies to manage traffic on freeways and arterials between downtown Nashville and the city of Murfreesboro. The freeway segment experiences an \textit{Annual Average Daily Traffic} (AADT) in excess of 160,000 vehicles, with peak hours marked by significant congestion and frequent stop-and-go patterns~\cite{gloudemans202324}. 
% Serving as a crucial path for both commuter and freight traffic, including 10-15\% heavy trucks, I-24 plays a vital role in facilitating shipping and industrial transportation within Middle Tennessee and across the southeastern United States.

To improve traffic safety and travel time reliability, I-24 SMART Corridor integrates multiple \textit{Active Traffic Management} (ATM) strategies, including VSL, a lane control system, and arterial signal integration. Currently, I-24 SMART Corridor has deployed 34 VSL gantries on I-24 westbound and 33 VSL gantries on I-24 eastbound, spanning 17 miles from mile marker 53 to mile marker 70. In this area, 60 \textit{Radar Detection System} (RDS) sensors have been installed or upgraded to monitor traffic performance and provide state inputs to our MARL-based control algorithm at a 30-second interval, which is shorter than we assumed in training as the data rates changed as the system evolved. Figure~\ref{fig:smartcorridor} displays a map of the VSL deployment segment.

\begin{figure}
    \centering
    \includegraphics[width=1\columnwidth]{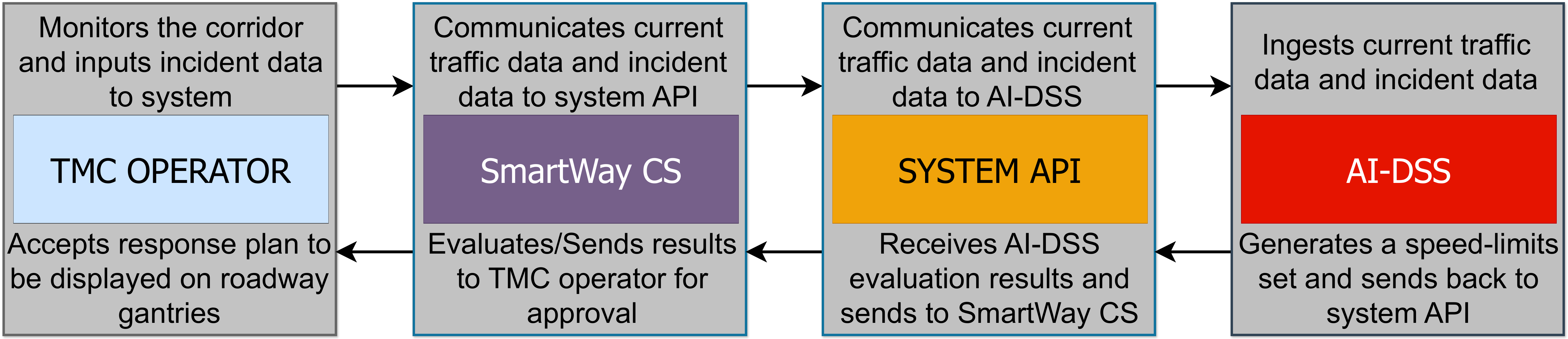}
    \caption{Overview of the  communication between AI-DSS, the TMC software SmartWay CS, and the TMC operator.}
    \label{fig:ai-dss work flow}
\end{figure}

\subsection{AI-DSS}

% 1. general work flow (system integration)
% 2. inter-system communication
% 3. development, testing, production.
To implement MARL-based controllers into the production active traffic management software (SmartWay CS) used in the regional \textit{Traffic Management Center} (TMC), we created a software stack known as the AI-DSS~\cite{van2022system}. Figure~\ref{fig:ai-dss work flow} presents the workflow of the communications between AI-DSS and the TMC. The TMC operator monitors the corridor conditions and records relevant incident information in SmartWay CS. An API in SmartWay CS allows bidirectional communications with the AI-DSS over the TCP/IP protocol using websockets. Based on the real-time traffic information from SmartWay CS, the AI-DSS implements the MARL-based control algorithm and provides the speed limits to be posted back to SmartWay CS. SmartWay CS verifies that the speed limits do not violate any constraints, and posts the speed limits to the gantries on the roadway.

The AI-DSS is implemented in Python for its extensive support for libraries enabling multi-processing, websocket connectivity, database logging, and the execution of MARL-based policies. Currently, five separate environments are designated for the AI-DSS: development, testing, production mirror at Vanderbilt, demo, and production at Tennessee Department of Transportation (TDOT). The development environment is utilized for debugging with real-time data. After comprehensive testing of the AI-DSS with the MARL-based VSL control algorithm, it is deployed to the TDOT production environment for real-time traffic control.

% \Zhiyao{it seems more like internal debugging/monitoring config}
% \Zhiyao{A general feedback is that the way we introduce AIDSS should be more on how it `supports' the deployment but not how it is designed (and it's not open-source, so it doesn't really matter why we use Python though it's good to mention). It may be better if we focus on its roles and enumerate/itemize its functionalities, e.g.: (1) establishes bidirectional communication to TMC through API, (2) comprehensively logs related information into the database for system monitoring, (3) provides a highly modular system architecture which is compatible with additional functionalities and lowering the bar to implement in other locations, etc.}

%%%%%%%%%%%%%%%%%%%%%%%%%%%%%%%%%%%%%%%%%%%%%%%%%%%%%%%%%%%%%%%%%%%%%%%%%%%%%%%%
\section{Results}\label{sec:results}
In this section, we present the MARL-based VSL control algorithm deployment results. First, we show the control algorithm behavior from a random morning peak hour. Next, we display the effective control time of the MARL-based policy (with IAM) and the safety guards in the algorithm. Finally, we quantify the domain mismatch between the simulation and real-world observations and demonstrate the robustness of the learned policy.
% \Zhiyao{``Component effectiveness percentage" is a bit confusing. I cannot figure out a super appropriate term. @Dan}

% 1. qalitative results of VSL behavior
% 2. quantitative results for effective percentage of each component
% 3. quantitative results to show the wasserstein distance among datasets.
\subsection{Algorithm Behavior}
% \Yuhang{A name for the deployed algorithm?}
Figure~\ref{fig:control algorithm behavior} (a) shows the time-space diagram of the average traffic speed of the morning peak hour of I-24 Westbound on Monday, April 22, 2024. The x-axis represents time and y-axis represents mile markers of the 17-mile segment of I-24, where the traffic is going upward along y-axis to downtown Nashville. With colors denoting the traffic speed recorded by RDS sensors, Figure~\ref{fig:control algorithm behavior} (a) exhibits a typical morning rush hour congestion pattern of the selected I-24 segment, with the first congestion wave occurring at 5:30 AM.
% and the final congestion clearing after 10 AM.

Figure~\ref{fig:control algorithm behavior} (b) displays the time-space diagram for 34 VSL gantries on I-24 Westbound, which are controlled by the MARL-based VSL algorithm described in Section~\ref{sec:methods} at 30-second intervals, with the same time and space ranges as shown in Figure~\ref{fig:control algorithm behavior} (a). Note that there are 6 consecutive VSLs closest to downtown Nashville with a smaller maximum speed limits than the rest of the VSLs, as determined by TDOT to improve traffic safety. To take a closer look at the role of MARL-based policy (with IAM) in our control algorithm, Figure~\ref{fig:control algorithm behavior} (c) presents the same diagram as Figure~\ref{fig:control algorithm behavior} (b) but with all safety guards masked as white.

The behavior of the algorithm can be described based on three different traffic regimes: congestion regime, free-flow regime, and transition regime. As shown in Figure~\ref{fig:control algorithm behavior} (a), the congestion regime can be identified as the dark red area, the free-flow regime as the dark green area, and the transition regime as the alternating yellow, orange and shallow green area. Specifically, the MARL-based policy (with IAM) is able to identify the congestion and free-flow regimes for most times due to the adaptability and mobility reward terms and the informative state space design. As for transition regime, we can divide it into three categories, namely, \textit{free-flow to congestion} (F-C), \textit{congestion to congestion} (C-C), and \textit{congestion to free-flow} (C-F). With a comparison between Figure~\ref{fig:control algorithm behavior} (c) and Figure~\ref{fig:control algorithm behavior} (a), we can observe that the MARL-based policy (with IAM) can generate a smooth slow-down speed profile for F-C thanks to the safety reward term and the involvement of the invalid action masking. However, we still need Speed-Matching and Debounce safety guards to work for C-C and C-F with a goal to satisfy the authority requirements. Finally, we note that the white part on the top of Figure~\ref{fig:control algorithm behavior} (c) is because of the 6 VSLs with smaller maximum speed limits, for which the Maximum Speed Limit Correction safety guard has been triggered. 

Finally, to better understand how the deployed algorithm behaves from a driver's perspective, we generate the trajectories of 3 simulated vehicles according to the RDS speed data from Figure~\ref{fig:control algorithm behavior} (a). Figure~\ref{fig:control algorithm behavior} (d) shows the time series of travel speed and the corresponding speed limits for each simulated vehicle. Starting from 6 AM, Vehicle 1 encounters multiple stop-and-go waves along its journey. Meanwhile, the VSL can inform Vehicle 1 of the incoming slow-down or speed-up patterns in advance for most times, as we can observe a time lag between the blue-dashed line and the orange line in Figure~\ref{fig:control algorithm behavior} (d). With a later starting time and a longer travel time, Vehicle 2 and Vehicle 3 encounter a stand-still congestion pattern, during which the VSL behaves in advance to provide the slow-down warning signal for the vehicles, aiming to prevent a sudden break or at least inform the upstream drivers of the downstream traffic condition.

\begin{table}
\caption{The daily effectiveness percentage (AVG$\pm$STD) of MARL-Policy with IAM (Policy), Speed-Matching (SM), Maximum Speed Limit Correction (MSLC), and Debounce (DB). Note ``I'' and ``E'' refer to including gantries with custom max speed limit and excluding gantries with custom max speed limit, ``WB'' and ``EB'' refer to ``Westbound'' and ``Eastbound'', ``PH'' refers to ``Peak Hour''.} 
% \Zhiyao{table caption is hard to read. move acronym definitions to main content}
  \centering
  \begin{tabularx}{\columnwidth}{@{}llcccc@{}} % "X" indicates a column that will expand to fill the text width
    \toprule
    & Dataset     & Policy (\%)  & SM (\%) & MSLC (\%) & DB (\%) \\ 
    \midrule
    \multirow{4}{*}{\begin{tabular}[c]{@{}l@{}}I\end{tabular}} 
    & I-24 WB      &81.3$\pm$0.8    &1.8$\pm$1.1      &16.1$\pm$1.1  &0.8$\pm$0.5    \\
    & I-24 WB PH &78.3$\pm$2.1    &7.4$\pm$1.3     &10.5$\pm$2.1  &3.8$\pm$0.5  \\
    & I-24 EB      &87.3$\pm$0.9    &2.8$\pm1.6$      &9.6$\pm$1.2   &0.3$\pm$0.2 \\
    & I-24 EB PH &84.3$\pm$2.9    &12.8$\pm$2.7      &1.2$\pm$1.2   &1.7$\pm$0.6 \\ 
    \midrule
    \multirow{4}{*}{\begin{tabular}[c]{@{}l@{}}E\end{tabular}} 
    & I-24 WB      &98.4$\pm$1.0    &1.3$\pm$0.8      &0    &0.3$\pm$0.2\\
    & I-24 WB PH   &93.1$\pm$1.5    &5.1$\pm$1.1      &0    &1.8$\pm$0.4\\
    & I-24 EB      &97.5$\pm$1.7    &2.2$\pm$1.5      &0    &0.3$\pm$0.2\\
    & I-24 EB PH   &86.6$\pm$3.7    &11.8$\pm$3.2      &0    &1.6$\pm$0.7\\ 
    \bottomrule
  \end{tabularx}
\label{table:percentage}
\end{table}

\subsection{Control Effectiveness Analysis}
% In Section~\ref{sec:methods}, we discussed the four components of the deployed VSL control algorithm, i.e., MARL-based policy (Policy), Speed-Matching (SM), Maximum Speed Limit Correction (MSLC) and Debounce (DB). 
Given a dataset collected from March 8, 2024 to April 24, 2024 with 8,923,106 decisions for 67 gantries, Table~\ref{table:percentage} displays the amount of time the MARL-based policy with IAM (Policy) is implemented directly, the amount of time that Speed-Matching (SM) is used to correct the Policy and the amount of time that Maximum Speed Limit Correction (MSLC) and Debounce (DB) is used for final adjustments.

On average, the Policy has controlled 81.3\% of the time on I-24 Westbound and 87.3\% on I-24 Eastbound daily, across all 67 VSL gantries. To further understand the situation during peak hours, we analyzed the morning peak hour on I-24 Westbound and the afternoon peak hour on I-24 Eastbound, during which the Policy has a slightly reduced controlled time. It is worth to note that the gantries with a customized max speed limit will trigger MSLC constantly when traffic is in freeflow. We remove the 10 gantries with a custom max speed limit, i.e., 6 from Westbound and 4 from Eastbound, and show the results in the bottom part of Table~\ref{table:percentage}. Among those 57 gantries with the same max speed limit of 70 mph, the Policy has controlled 93.1\% of the time for westbound morning peak hour and 86.6\% of the time for eastbound afternoon peak hour. 

\begin{figure}
    \centering
    \includegraphics[width=1\columnwidth]{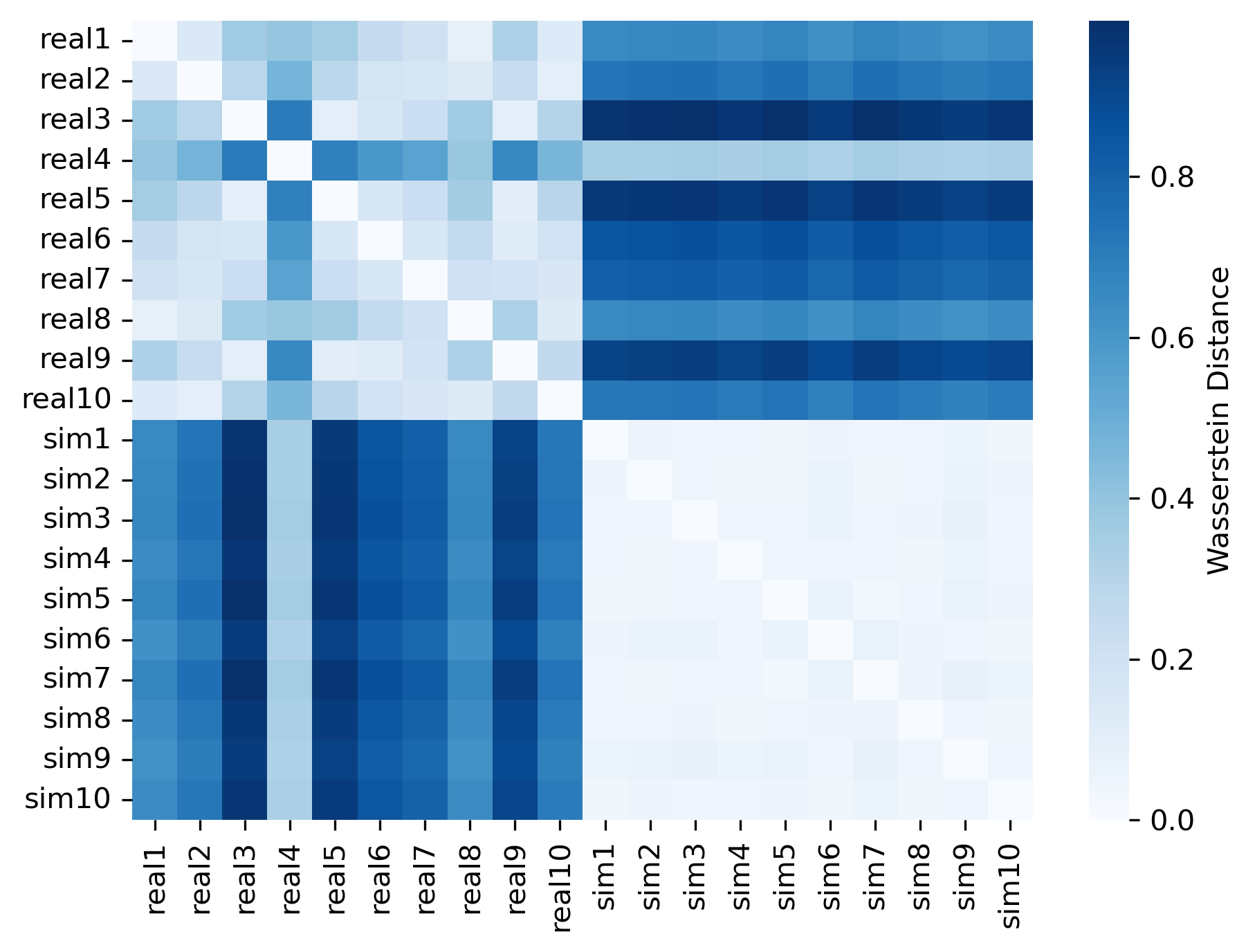}
    \caption{The Wasserstein Distance across the real-world and simulation observations.}
    \label{fig:wasserstein distance}
\end{figure}
\subsection{Domain Mismatch Quantification}
% It is widely acknowledged that a significant obstacle to deploying RL in real-world settings is the so-called sim-to-real gap. There are mainly two categories of the gap: \textit{domain mismatch} and \textit{model mismatch}. Domain mismatch refers to the situation where   
One of the challenges that impede the successful transfer of simulation-based RL policies to real-world deployment is \textit{domain mismatch}, where the real-world observations may not overlap with the ones in simulation. To quantify this mismatch, we calculate the Wasserstein Distance~\cite{flamary2021pot} among the observation samples between simulation and real world. 
% \Zhiyao{Add one or two sentences to describe the WD. Saying WD is a measure of two probabilistic distributions in the sense of how ``costly" one transforms to the other. Then give the math definition. Then say what are the distributions in our experiment. Then it should be adhesive to the rest content}
In detail, we conduct 10 random experiments in the simulation testing environment as described in Section~\ref{sec:methods}. We then randomly sample 1000 observation data points of each experiment to generate 10 datasets on the simulation side, i.e., sim1 to sim10. Similarly, we sample 1000 observation samples of I-24 Westbound peak hour from each of the 10 random days to generate real-world datasets.

Figure~\ref{fig:wasserstein distance} presents the heatmap of the Wasserstein Distance across every two pairs of the aforementioned 20 datasets. With a symmetric structure, this heatmap demonstrates three facts. First, the observation distributions of different simulation experiments are very close to each other as can be seen from the bottom right part of Figure~\ref{fig:wasserstein distance}. Second, the observation distribution of real world has a shift from day to day, as seen from top left of Figure~\ref{fig:wasserstein distance}, indicating a varying traffic pattern with different days. Third, the distance between real-world and simulation observations is larger than that within real-world datasets or simulation datasets. With this domain mismatch, the learned MARL-based policy (with IAM) demonstrates a robust performance, as can be seen from Table~\ref{table:percentage}.

% FIGURE 5 todo: make the y-axis consistent(make them the same), the color bar consistent, maybe we can add the difference between these two heatmaps, maybe we can add the time-series of the virtual vehicles

% Figure 6 & 7, make them as a table

% Figure 8, change real1 to day1..., maybe we can add a time-space diagram of both the simulation and real-world to show their difference.

% Add a figure for the overall work flow

% we train a marl on a real-freewayk as the system continue to run we will do teiled annalyiss with respect to the traffic satey and impact and other fators, here we only provide the preliminary results of the MARL algorithm that has been running currently.

%%%%%%%%%%%%%%%%%%%%%%%%%%%%%%%%%%%%%%%%%%%%%%%%%%%%%%%%%%%%%%%%%%%%%%%%%%%%%%%%
\section{Conclusions}\label{sec:conclusions}
This work describes the first MARL deployment of a VSL control system on the I-24 freeway in Nashville, TN, which  continues to operate today. The preliminary results demonstrate that it is possible to deploy a simulation-based MARL policy in the real world with safety guards. The safety guards are needed but run only a small portion of time compared to the RL policy, demonstrating the potential for further RL-based deployments on infrastructure systems. As the system continues to run, we expect to be able to provide more datasets and analysis of the performance of the algorithm with respect to traditional safety and performance measures on the corridor. 

%%%%%%%%%%%%%%%%%%%%%%%%%%%%%%%%%%%%%%%%%%%%%%%%%%%%%%%%%%%%%%%%%%%%%%%%%%%%%%%%

%%%%%%%%%%%%%%%%%%%%%%%%%%%%%%%%%%%%%%%%%%%%%%%%%%%%%%%%%%%%%%%%%%%%%%%%%%%%%%%%



% \section*{ACKNOWLEDGMENT}

% The preferred spelling of the word ÒacknowledgmentÓ in America is without an ÒeÓ after the ÒgÓ. Avoid the stilted expression, ÒOne of us (R. B. G.) thanks . . .Ó  Instead, try ÒR. B. G. thanksÓ. Put sponsor acknowledgments in the unnumbered footnote on the first page.

%%%%%%%%%%%%%%%%%%%%%%%%%%%%%%%%%%%%%%%%%%%%%%%%%%%%%%%%%%%%%%%%%%%%%%%%%%%%%%%%

% References are important to the reader; therefore, each citation must be complete and correct. If at all possible, references should be commonly available publications.

\bibliographystyle{ieeetr}
\bibliography{root}

\end{document}